# The Activation Entropy Change in Enzymatic Reaction Catalyzed by Isochorismate-Pyruvate Lyase of Pseudomonas Aeruginosa PchB

Liangxu Xie, Zhe-Ning Chen, Mingjun Yang

Department of Chemistry, The University of Hong Kong, Pokfulam Road, Hong Kong S. A. R., China

# ABSTRACT

The elucidation of entropic contribution to enzyme catalysis has been debated over decades. The recent experimentally measured activation enthalpy and entropy, for chorismate rearrangement reaction in PchB brings up a hotly debated issue whether the chorismate mutase catalyzed reaction is entropy-driven reaction. Extensive configurational sampling combined with quantum mechanics/molecular mechanics molecular dynamics (QM/MM MD) provides an approach to calculate entropic contribution in condensed phase reactions. Complete reaction pathway is exploited by QM/MM MD simulations at DFT and SCC-DFTB levels. The overall entropy change −13.9 cal/mol/K calculated at SCC-DFTB level QM/MM MD simulations, is close agreement with the experimental value −12.1 cal/mol/K. Conformation analysis indicates that the self-ordering of chorismate in the active site of PchB also contributes to total entropy change. This entropy penalty including conformational transformation entropy and activation entropy cannot be intuitively speculated from the crystal structure that only acts as a stationary state along the reaction pathway of PchB catalyzed reaction. This is the first time to use QM/MM MD simulations to calculate the activation entropy from the temperature dependence of reliable free energy profiles with extensive simulation time. The reasonable insight in enthalpy/entropy scheme clarifies the detailed entropy change and provides a quantitative tool to the contradicted experimental results.

**Abbreviations:** CM, Chorismate mutaste; US, umbrella sampling; TS, transition state; RS, reactant state; RC, reaction coordinate; QM/MM MD, quantum mechanics/molecular mechanics molecular dynamics; PchB, Isochorismate-Pyruvate Lyase of Pseudomonas aeruginosa; NMR, Nuclear magnetic resonance (NMR); RMSD, root mean square deviation; RMSF, root mean square fluctuation.

# INTRODUCTION

The role of entropy to catalysis has been widely investigated. Commonly the entropy is interpreted in terms of a loss of translational and rotational motions of substrate when it passes from reactant to transition state (TS). In water, the entropy loss is from either the ordering of reactant or reorganizing of water. Correspondingly, enzyme either pays entropy penalty by ordering substrate or pre-organizing active site in comparison with reaction in water. Enzyme can bind their substrate tightly before catalyzed reaction. It has been accepted that enzyme has already paid for the translational and rotational entropy during binding to achieve high efficiency as pointed by Page and Jencks[1-2]. The activation barrier is mainly reduced from favorable entropy change. However, Bruice and Warshel have shown that activation barrier was dominated by reduction of enthalpy $\Delta H^{\ddagger}$ rather than entropy $-T\Delta S^{\ddagger}$.[3-4] The key contribution of catalysis is actually the transition state stabilization (TSS)[4-6], in which *TS* is stabilized by electrostatic interaction from active site residues[4, 7]. Dominant electrostatic interactions imply that enthalpy contribution is critical to the CM catalysis. These arguments make it attractive to directly delineate the magnitude of entropy contributes to the catalysis by quantitative method. Previous theoretical studies have exclusively focused on the entropy of substrate. In condensed phase, huge number of degrees of freedom is involved in the enzymatic reactions. Åqvist has made great progress in computer calculation of activation entropy by including contributions from solvent and protein which have been dismissed in previous reports[8-10]. From extensive MD simulations, Åqvist found that the calculated thermodynamic activation parameters are in remarkable agreement with experimental results.[8] The computed activation entropy scheme proposed by Åqvist makes it possible to directly compute activation entropy from simulations.

The rearrangement of chorismate to prephenate catalyzed by chorismate mutases (CMs), a key step in the biosynthesis of phenylalanine and tyrosine, is a widely investigated enzymatic reaction (Scheme 1). In addition to CMs catalyzed reaction, "isochorismate–pyruvate lyase from *Pseudomonas aeruginosa*", designated as PchB, also shows chorismate mutase activity. Different from the traditional, high-specificity functional enzymes, PchB shows capability of catalyzing the primary reaction of conversion isochorismate into pyruvate, and catalyzing secondary reaction of transforming chorismate into prephenate as described in Scheme 1.[11-14] Promiscuous enzymes display more conformational flexibility than their more evolved, highly selective counterparts.[15-17] This promiscuous activity has attracted experimental and theoretical studies to improve its secondary activity by mutations.[18-20] Clarifying the entropic penalty thus becomes necessary to improve its secondary activity of PchB. However, the recently experimental activation enthalpy and entropy, for chorismate rearrangement reaction in PchB,[21] invokes much debate. No consistent conclusion has been made that how much the entropic contribution to catalysis because the measured entropy takes up different proportion in the reduced energy barrier for different CMs catalyzed reactions.[21-24]

The controversy surrounding the relative importance of enthalpy and entropy in PchB catalysis has not yet been resolved, despite available experimental and theoretical studies.[20-21] Structural analysis indicates that most of important catalytic residues of PchB are identical to those of EcCM, such as two highly charged arginine residues[25]. Thermodynamic parameters obtained in recent kinetic experiments state that EcCM cause less entropy change,[22] while PchB needs a relative large entropy change.[21] The crystal structure cannot provide enough clues to explain the large measured entropy change.[21, 26] Two carboxylate groups are oriented by two arginine residues in the active sites of PchB in the same manner to that of EcCM[26]. It is expected that the extra conformational constriction on substrate should already reduce the entropy in the Michaelis complex. Luo and coworkers provided one hypothesis that a reactive conformation of the substrate may not be formed upon binding of substrate and thus the entropy penalty comes from the ordering of substrate and loop upon the reaction coordinate approaching the transition state.[21] This is consistent with the idea that enzyme dynamic motions promote catalysis.[27-28] The conformational change of chorismate from unreactive state to reactive state before catalyzed reaction has been found in both yeast CM and BsCM from QM/MM MD simulations. After binding the prevalent nonreactive substrate, enzymes can transform them into the reactive state in chair-like pseudodiaxial conformation before the chemical reaction.[29-30] Martí also reported that chorismate underwent a nonreactive to reactive state in the BsCM from a preorganization view.[31] To calculate the entropy change, we carried out QM/MM MD simulations at SCC-DFTB level.

In this study we focus on the enthalpy and entropy change of the PchB-catalyzed reaction. Decomposition of the free energy into enthalpy and entropy contribution was used to illustrate thermodynamic contribution ($\Delta H^\ddagger$) to *TS* stabilization and activation entropy ($\Delta S^\ddagger$) from energetic view. As stated by Warshel, simulations for enzyme catalyzed reaction still lacks complete configurational sampling to confirm the reaction mechanism using QM/MM approaches.[4] From recent work of Åqvist, microsecond simulations using EVB method can be used to get remarkably accurate entropy value[8-9]. Extensive sampling is important to guarantee the accuracy of entropy calculation. Validation this scheme using QM/MM method without optimizing parameters has not been performed so far because of huge computational cost. Herein, we use SCC-DFTB method to calculate the activation entropy contribution at 4.9 μs timescale (close to Åqvist's simulation time 10 μs).

In this contribution, we first discuss conformations of chorismate in water and in PchB; then the activation energy change of the reaction and structural details are presented. The entropy changes in both conformation change process ($\Delta S_c$) and reaction process ($\Delta S^\ddagger$) are computed using QM/MM MD simulations.

## RESULTS

**Chorismate Conformational Change**

Nuclear magnetic resonance (NMR) studies have shown two dominant conformations of chorismate in solution, pseudo-diequatorial and pseudo-diaxial conformations.[32] The equilibrium among different states at different semi-empirical quantum levels has been extensively discussed in previous reports.[33-34] The dihedral C3–O7–C8–C9 and dihedral C2–C3–O7–C8 have been used to distinguish different conformations. Figure 1 shows the dihedral changes during QM/MM MD simulations when chorismate is treated with SCC-DFTB level.[35-36] The representative structures and free energy landscape of chorismate in water are shown in Figure S1, S2 and Table S1. In PchB, no transition has been observed for the binding chorismate in the 10 ns QM/MM MD simulation.

This conformational change of chorismate after binding actually has been reported in both yeast CM and BsCM, whcih can bind the prevalent nonreactive substrate in pseudodiaxial conformation, and transform them into the reactive state in chair-like pseudodiaxial conformation before the chemical reaction.[29-31] The same reaction coordinate has been used to show this conformational change in PchB as shown in Figure S3.

By scanning the free energy change along the C3–O7–C8–C9 dihedral, two minima are determined from the PMF plot for the substrate in water and in PchB as shown in Figure 2A. The global minimum in water and in the enzyme has been shifted at the same energy level for comparison. In water, transformation of the C3–O7–C8–C9 dihedral from the second minimum to the global minimum needs a lower barrier, 3.1 kcal/mol. In PchB, there is one local minimum (state *R1*, uncompressed chorismate) and one global minimum (state *R2*, compressed chorismate). State *R1* of PchB is close to state *S5* in solution (Figure S2). The transformation from local to global state shows a low energy barrier (0.49 kcal/mol) at $148^0$ in PchB.

PMF profile in Figure 2A implies that PchB binds unreactive conformers. To determine whether pre-orientation of chorismate occurs spontaneously after binding, one conformation around this local minimum is selected as a starting conformation to continue a 2000 ps QM/MM MD simulation. Figure 2B shows that conversion from local to global minimum is automatic (conversion occurs at 1400 ps). This fast transformation process may be hard to be determined by NMR and ignored in experiments. After crossing the barrier, the energy of the global minimum is 4.7 kcal/mol lower than the second minimum state, and the substrate stays at the minimum state ($-91^0$). Both the local minimum and global minimum of chorismate/PchB complex show relative stability. Arg31 forms strong salt bridge with one of carboxylate group of chorismate. Lys42 and Gln90 establish hydrogen bonds with O7 of the ether group in chorismate. When transforming from local state to global state, the carboxylate group of the side chain rotates, forming hydrogen bonds with Arg14(B) (Arg14 from chain B). In the active site, the ring of chorismate rotates in a slight angle to adopt its new position as described in Figure 2C and 2D at the meantime the enzyme will keep its structure revealed from RMSD in Figure S4. Arg31, Arg14 (from chain B) and Lys42 together contribute to form reactive state of chorismate in PchB, as revealed from MD representative structures. The chorismate is compressed into an ordered state in the active site of the PchB due to restricted space and favorable interaction.

Similar to yeast CM and BsCM,[29-30] PchB can bind unreactive chorismate in solution. To confirm this idea, the whole reaction process has been estimated using short MD simulations with simulation time of 100 ps for each umbrella windows. The PMF profiles in Figure 3 shows that the existence of state *R1* and *R2* is important to construct the whole process. PchB transforms uncompressed state into reactive state in a following step of enzymatic reaction. This step may contribute entropy change, which has been ignored.

For this processes (R1→R2), the entropy can be estimated using Quasi-harmonic analysis (QHA) and Schlitter formula, which has been extensively examined.[37-39] The entropy calculated from Schlitter's formula [38] provides an alternative way to estimate entropy despite of overestimation of entropy change.[51] The estimated configurational entropy is −10.5 cal/mol/K for preorganization step when chorismate changes from uncompressed state to compressed state in the active site of PchB.

**Free Energy of Reaction Catalyzed by PchB**

The free energy calculation has been carried out by using B3LYP/6-31G(d) to treat the QM region. PMF profiles for the reaction in water and in PchB are shown in Figure 4. The activation energy barriers $\Delta G^{\ddagger}$ calculated by DFT are 22.0 and 17.3 kcal/mol for the uncatalyzed reaction and for the enzymatic reaction, respectively, at 298 K. Energy barriers calculated by DFT are 2.5 and 2.2 kcal/mol lower than experimental uncatalyzed reaction and catalyzed reaction, respectively.

This DFT result is consistent with experimental data, considering that DFT normally underestimates free energy barriers by 3–5 kcal/mol because of poor treatment of dispersion interactions.[40] Underestimation of the activation energy barrier by ~3 kcal/mol with DFT in other CM systems also has been reported by Claeyssens and coworkers.[41] PchB reduces the free energy barrier by 4.7 kcal/mol in current calculation ($\Delta G^{\ddagger}_{uncat}$- $\Delta G^{\ddagger}_{cat}$), which is very close to the experimental result, 4.97 kcal/mol. The B3LYP/6-31G(d)/CHARMM level of the QM/MM theory thus gives an accurate energy barrier.

To evaluate entropy changes quantitatively, long-time simulations are required to explore the conformational space. Considering both the accuracy and computational cost, SCC-DFTB is adopted in the following simulations. Potential of mean force (PMF) plots for the enzymatic and uncatalyzed reactions from 20 replicas for US at one temperature are shown in Figure 4. The reaction energy changes $\Delta G$ of the enzymatic reaction at the DFT level (−13.4 kcal/mol) is similar to that at the SCC-DFTB level (−13.8 kcal/mol) around the product state while SCC-DFTB method estimates the energy around transition states. The activation barrier $\Delta G^{\ddagger}$ for uncatalyzed reaction is 15.6 ± 0.2 kcal/mol, and the barrier for enzymatic reaction is 12.1 ± 0.1 kcal/mol (Table 1). The lower activation energy barrier than experimental value is due to the electron tight binding in the SCC-DFTB method for this system.[42] The statistical error calculated from multiple simulations is below 2% among 20 replicas in SCC-DFTB/MM simulations (Table 1). The enzymatic reaction shows free activation energy 3.5 kcal/mol lower than that of the uncatalyzed reaction. This free energy difference is closer to the experimental value than a previous reported value at the AM1 level (3.5 kcal/mol vs. 2.3 kcal/mol)[20].

**Activation Entropy**

The enthalpy change is thermodynamics contribution of the *TS* stabilization, while entropy change is the thermodynamics contribution from dynamics fluctuation when Michaelis complex converts to TS/enzyme complex. To calculate activation entropy, the convergence of error associated with $\Delta G^{\ddagger}$ should be less than 0.2 kcal/mol [8, 10] and this requirement has been achieved within 500 ps as shown in Figure S5. Decomposition of the total reduced 3.5 kcal/mol into enthalpy and entropy contributions provides a clearer description of the interaction and dynamic fluctuation changes during reaction. On the basis of the linear fitting, the entropy values are −8.6 and −3.4 cal/mol/K for the uncatalyzed reaction and enzymatic reactions, respectively (Figure 5). Note that high adjusted R-square (equal or slight larger than 0.97) of linear fitting proves the reliable of current entropy calculation approaches from fitting with Gibbs-Helmholtz equation. The activation entropy is derived from free energy difference at different temperature and the systematic error can be removed.[43] Clearly, the activation entropy indicates that the activation entropy of the enzymatic reaction is smaller than that of the corresponding reaction in solution. This is consistent with a previous hypothesis.[21] The total reduced 3.5 kcal/mol free energy barrier ($\Delta\Delta G^{\ddagger}$) can be decomposed into 2.1 kcal/mol for the enthalpy contribution ($\Delta\Delta H^{\ddagger}$) and 1.4 kcal/mol for the entropy contribution ($-\Delta(T\Delta S^{\ddagger})$).

The conformation change from *R2* to *TS* in PchB is smaller than that of the uncatalyzed reaction. The average distance and dihedral of the *TS* and *RS* are analyzed and summarized in Table 2. A significant difference in conformation changes exists between the reactions in water and in PchB. The C3–O7 distances (bond breakage) in the *RS* are 1.49 and 1.51 Å in water and in the enzyme, respectively. Hydrogen-bonding or electrostatic interaction between O7 and Gln90 and Lys42 can stabilize the breaking C3–O7 bond. In contrast, the C1–C9 distance (bond formation) is shorter in the enzyme than in water. The two carboxylate C10–C11 distance is shorter in the enzyme because of stabilization by arginine residues, Arg14 and Arg31. Contrast, the two carboxylate groups move as far as possible in water because of electrostatic repulsion. Since the C3–O7–C8–C9 dihedral in the enzyme is distinct from that in water, C9 does not achieve the right orientation toward C1 in water, which is attained in the enzyme. The C3–O7–C8–C9 dihedral shows large changes in the uncatalyzed reaction during chemical reaction process. In water, chorismate shows more flexibility in the *RS* without the surrounding strained force. The entropy penalty is dominated by ordering the reactant.

For reaction process, the decreased enthalpy (2.1 kcal/mol, accounts for 60% of total reduced energy barrier) shows slightly larger contribution than entropy contribution. PchB tends to stabilize *TS* through electrostatic interaction. The shortened distance between O7 and Lys42 when chorismate changes from *RS* to *TS* displays the more favorable electrostatic interaction for binding *TS* (Table 3). Electrostatic interactions between chorismate and enzyme can be verified from the distance between heavy atoms of residues of the enzyme and substrate from the opposite atomic charges (Schematic figure is shown in Figure S6). Selected distances for chorismate and the closest residues of the enzyme are summarized in Table 3. An obvious difference between *RS* and *TS* is the decrease in distance between ether oxygen (O7) and neighboring residues. In the *TS*, the broken C3–O7 bond is stabilized by stronger electrostatic interactions. The O7–NZ~Lys42 distance decreases from 3.12 to 2.99 Å, as well as the O7–N~Gln90 distance decreases from 2.98 to 2.92 Å. Lys42 and Gln90 establish a strong hydrogen bond and electrostatic stabilization with ether oxygen O7. Among them, O7–NZ~Lys42 shows the largest distance change when *RS* converts into *TS*. Therefore, electrostatic interactions between Lys42 and chorismate during chemical conversion contribute to the catalysis.

## DISCUSSION

Herein conformation transformation step and chemical reaction step are combined to understand the reaction mechanism in PchB as shown in Figure 6. Reactant compression occurs before Claisen rearrangement reaction in the active site. PchB shares some similarity with BsCM and EcCM and meantime shows its unique features. PchB binds uncompressed chorismate and convert it to compressed state. The compressed state is stabilized by favorable interactions between chorismate and neighboring residues. The crystal structure represents one of stable states and cannot be used to infer the whole process of enzymatic reactions. For PchB, the existence of state *R1* indicates that ordering of reactant cannot be ignored. The conformation transformation has been used to explain the catalytic mechanism. However, no connection has been built to explain the entropy change of PchB catalyzed reaction. The ordering of substrate contributes to −10.5 cal/mol/K entropy change as estimated from QHA and Schilitter formula (Table S3). This ordering step may be coupled with reaction step. The reaction step is the rate-limiting step, and the free energy change during the reaction step is 12.1 kcal/mol. The activation entropy is −3.4 cal/mol/K calculated in a rigorous way. The activation entropy is similar to that of EcCM that shares similar structural topology. This smaller entropy changes is consistent with conformational changes from compressed reactant to *TS*. The total entropy change in transformation process and in reaction process is −13.9 cal/mol/K, which is close to experimental value, −12.1 cal/mol/K. The overestimation is from QHA method which provides the upper limit of entropy change. The result is capable of elucidating the whole reaction process. For large entropy process, the entropy can be estimated from QHA or Schilitter formula considering the convergence and computational cost. For activation entropy process, entropy can be calculated from converged free energy changes. The agreement of calculated entropy with experimentally measured entropy illustrates that the measured entropy is actually the apparent entropy change. The measured entropy covers entropy from both ordering step and chemical reaction step.

## CONCLUSIONS

QM/MM MD simulations have carried out for the conformation transformation process and chemical reaction process to illustrate ambiguous entropy change in PchB-catalyzed reaction. The convergence of free energy change is verified at DFT and SCC-DFTB level. By combining with QM/MM hybrid potential, MD simulations are capable of computing activation entropy change. PchB requires less entropy change in elementary chemical reaction step (−3.4 cal/mol/K in PchB vs. −8.6 cal/mol/K in water). The combined estimated configurational entropy and rigorously calculated activation entropy, in total −13.9 cal/mol/K, is in close agreement to experimental value (−12.1 cal/mol/K). It indicates that the experimental entropy should be equivalent to the apparent overall entropy change from uncompressed bound chorismate to *TS*, rather than the entropy from single reaction step, which clarifies debated entropy scheme in PchB. The available crystal structure actually stands for one of stable state of chorismate-bound PchB locating in the middle of binding process and reaction process.

## COMPUTATIONAL METHODS

X-ray structure of PchB with bound salicylate and pyruvate was selected as the starting structure (PDB ID: 3REM).[44] The original substrate was replaced by chorismate through structural alignment. The complex of enzyme and substrate was solvated by TIP3P water model in periodic box.[45] The size of the solvent box was $80 \times 80 \times 80$ Å. The chorismate was described using QM methods, while the rest of the systems was described by CHARMM27 force field with CMAP in this study.[46-47] The simulation for apo form without substrate followed the same procedure after removing the substrate. The uncatalyzed reaction in water was solvated in a small box size $40 \times 40 \times 40$ Å$^3$ with TIP3P water model. The long range electrostatic interaction was computed with the particle-mesh-ewald approach.[48-49] The short-range interaction was calculated with a cutoff 8.0 Å, and that of medium-range interaction was calculated within the range 8.0-12.0 Å. The equations of motion were updated using the multiple time step algorithm for short-range interaction (1 fs), medium-range interaction (4 fs) and long range-range interaction (8 fs), respectively.[50-51] The full system was minimized and then warmed up gradually to 298K using the NPT ensemble for 1 ns. The whole system was further equilibrated using NVT ensemble in 1 ns MD simulations, and the temperature of system is maintained at 298K with Berendsen thermostat with relaxation time of 0.1 ps.[52] All simulations were performed using an in-house QM$^4$D software package.[53]

To verify further the conformational change in bound chorismate in active site of PchB, the C3–O7–C8–C9 dihedral was selected as the RC. Thirty-six US windows with a same interval of $10^0$ were used to cover $-180^0$ to $180^0$ to drive conformational changes along the C3–O7–C8–C9 dihedral. The simulation time for each window in PchB was 10 ns, while it was 1 ns in water. PMF can be constructed using weighted histogram analysis method (WHAM).[54-55]

The free energy calculation was performed by using QM/MM MD approach. To calculate free energy changes, the chorismate was treated by B3LYP/6-31G* basis set, while the rest was treated by CHARMM27 force field with CMAP in this study.[46-47] US method is expected to provide more accurate and reliable free energy.[56] This reaction coordinate is described by the antisymmetric combination of the C3–O7 and C1–C9 distances (denoted as RC). Simulation of the reaction was performed by using harmonic restraint at RC values ranging from −2.4 to 2.4 Å, with a step size of 0.1 Å. In each US window, a 20 ps production run was carried out for QM/MM MD simulation at the DFT level.

To further decompose the free energy into activation enthalpy and activation entropy, long timescale simulation is needed for sufficient sampling over the configuration space. To balance between the efficiency and accuracy, the simulation time of each window last 1 ns for the QM/MM MD simulation at the SCC-DFTB level.[35-36] Twenty independent simulations have been utilized for each window to estimate statistical average. The total simulation time for 49 windows was summed to 980 ns at one temperature. Free energies at five temperatures are introduced for a least–squares fitting method. US simulations have been performed at 293, 298, 303, 308, and 313 K for the enzymatic reaction. Relative smaller temperature interval is used to avoid enzyme conformational change.[57] For uncatalyzed reaction, the US simulations have been performed at 278, 288, 298, 308, and 318 K. Other 2D PMF profiles are computed from 2D umbrella sampling with 100 ps simulation time for each window.

**Additional Information**

Two dimensional free energy landscape of chorismate in water; representative structures of chorismate in water; key distance between residues and chorismate; entropy change of reaction.

# NOTES

The author declares no competing financial interest.

# ACKNOWLEDGEMENTS

The author is grateful to Dr. Hao Hu and Prof. Kwong-Yu Chan for insightful discussion. This research is conducted in part using the research computing facilities and/or advisory services offered by Information Technology Services, the University of Hong Kong.

# REFERENCES

1.      Jencks, W. P., Binding Energy, Specificity, and Enzymic Catalysis: The Circe Effect. In *Adv. Enzymol. Relat. Areas Mol. Biol.*, John Wiley & Sons, Inc.: 1975; pp 219-410.

2.      Page, M. I.; Jencks, W. P., Entropic Contributions to Rate Accelerations in Enzymic and Intramolecular Reactions and the Chelate Effect. *Proc. Natl. Acad. Sci. U.S.A.* **1971,** *68* (8), 1678-1683.

3.      Bruice, T. C.; Lightstone, F. C., Ground State and Transition State Contributions to the Rates of Intramolecular and Enzymatic Reactions. *Acc. Chem. Res.* **1999,** *32* (2), 127-136.

4.      Štrajbl, M.; Shurki, A.; Kato, M.; Warshel, A., Apparent NAC Effect in Chorismate Mutase Reflects Electrostatic Transition State Stabilization. *J. Am. Chem. Soc.* **2003,** *125* (34), 10228-10237.

5.      Shurki, A.; Štrajbl, M.; Villà, J.; Warshel, A., How Much Do Enzymes Really Gain by Restraining Their Reacting Fragments? *J. Am. Chem. Soc.* **2002,** *124* (15), 4097-4107.

6.      Barbany, M.; Gutiérrez-de-Terán, H.; Sanz, F.; Villà-Freixa, J.; Warshel, A., On the Generation of Catalytic Antibodies by Transition State Analogues. *ChemBioChem* **2003,** *4* (4), 277-285.

7.      Burschowsky, D.; van Eerde, A.; Ökvist, M.; Kienhöfer, A.; Kast, P.; Hilvert, D.; Krengel, U., Electrostatic transition state stabilization rather than reactant destabilization provides the chemical basis for efficient chorismate mutase catalysis. *Proc. Natl. Acad. Sci. U. S. A.* **2014,** *111*

(49), 17516-17521.

8. Kazemi, M.; Åqvist, J., Chemical reaction mechanisms in solution from brute force computational Arrhenius plots. *Nature Communications* **2015,** *6*, 7293.

9. kazemi shalkouhi, M.; Himo, F.; Aqvist, J., *Enzyme catalysis by entropy without Circe effect*. 2016; Vol. 113, p 201521020.

10. Åqvist, J.; Kazemi, M.; Isaksen, G. V.; Brandsdal, B. O., Entropy and Enzyme Catalysis. *Acc. Chem. Res.* **2017,** *50* (2), 199-207.

11. Gaille, C.; Kast, P.; Haas, D., Salicylate Biosynthesis in Pseudomonas aeruginosa PURIFICATION AND CHARACTERIZATION OF PchB, A NOVEL BIFUNCTIONAL ENZYME DISPLAYING ISOCHORISMATE PYRUVATE-LYASE AND CHORISMATE MUTASE ACTIVITIES. *J. Biol. Chem.* **2002,** *277* (24), 21768-21775.

12. DeClue, M. S.; Baldridge, K. K.; Künzler, D. E.; Kast, P.; Hilvert, D., Isochorismate Pyruvate Lyase:  A Pericyclic Reaction Mechanism? *J. Am. Chem. Soc.* **2005,** *127* (43), 15002-15003.

13. Künzler, D. E.; Sasso, S.; Gamper, M.; Hilvert, D.; Kast, P., Mechanistic insights into the isochorismate pyruvate lyase activity of the catalytically promiscuous PchB from combinatorial mutagenesis and selection. *J. Biol. Chem.* **2005,** *280*, 32827-34.

14. Luo, Q.; Olucha, J.; Lamb, A. L., Structure-function analyses of isochorismate-pyruvate lyase from Pseudomonas aeruginosa suggest differing catalytic mechanisms for the two pericyclic reactions of this bifunctional enzyme. *Biochemistry* **2009,** *48*, 5239-45.

15. Rosokha, S. V.; Kochi, J. K., The Preorganization Step in Organic Reaction Mechanisms. Charge-Transfer Complexes as Precursors to Electrophilic Aromatic Substitutions. *The Journal of Organic Chemistry* **2002,** *67* (6), 1727-1737.

16. James, L. C.; Tawfik, D. S., Conformational diversity and protein evolution – a 60-year-old hypothesis revisited. *Trends Biochem. Sci* **2003,** *28* (7), 361-368.

17. Hou, L.; Honaker, M. T.; Shireman, L. M.; Balogh, L. M.; Roberts, A. G.; Ng, K.-c.; Nath, A.; Atkins, W. M., Functional Promiscuity Correlates with Conformational Heterogeneity in A-class Glutathione S-Transferases. *J. Biol. Chem.* **2007,** *282* (32), 23264-23274.

18. Martí, S.; Andrés, J.; Silla, E.; Moliner, V.; Tuñón, I.; Bertrán, J., Computer-aided rational design of catalytic antibodies: The 1F7 case. *Angewandte Chemie (International ed. in English)* **2007,** *46*, 286-90.

19. Marti, S.; Andres, J.; Moliner, V.; Silla, E.; Tunon, I.; Bertran, J., Computational design of biological catalysts. *Chem. Soc. Rev.* **2008,** *37* (12), 2634-2643.

20. Martí, S.; Andrés, J.; Moliner, V.; Silla, E.; Tuñón, I.; Bertrán, J., Predicting an

Improvement of Secondary Catalytic Activity of Promiscuos Isochorismate Pyruvate Lyase by Computational Design. *J. Am. Chem. Soc.* **2008,** *130* (10), 2894-2895.

21.     Luo, Q.; Meneely, K. M.; Lamb, A. L., Entropic and Enthalpic Components of Catalysis in the Mutase and Lyase Activities of Pseudomonas aeruginosa PchB. *J. Am. Chem. Soc.* **2011,** *133* (18), 7229-7233.

22.     Galopin, C. C.; Zhang, S.; Wilson, D. B.; Ganem, B., On the mechanism of chorismate mutases: Clues from wild-type E. coli enzyme and a site-directed mutant related to yeast chorismate mutase. *Tetrahedron Lett.* **1996,** *37* (48), 8675-8678.

23.     Goerisch, H., On the mechanism of the chorismate mutase reaction. *Biochemistry* **1978,** *17* (18), 3700-3705.

24.     Kast, P.; Asif-Ullah, M.; Hilvert, D., Is chorismate mutase a prototypic entropy trap?- Activation parameters for the< i> Bacillus subtilis</i> enzyme. *Tetrahedron Lett.* **1996,** *37* (16), 2691-2694.

25.     Lamb, A. L., Pericyclic Reactions Catalyzed by Chorismate-Utilizing Enzymes. *Biochemistry* **2011,** *50* (35), 7476-7483.

26.     Zaitseva, J.; Lu, J.; Olechoski, K. L.; Lamb, A. L., Two crystal structures of the isochorismate pyruvate lyase from Pseudomonas aeruginosa. *J. Biol. Chem.* **2006,** *281*, 33441-9.

27.     Benkovic, S. J.; Hammes-Schiffer, S., A Perspective on Enzyme Catalysis. *Science* **2003,** *301* (5637), 1196-1202.

28.     Radkiewicz, J. L.; Brooks, C. L., Protein Dynamics in Enzymatic Catalysis:  Exploration of Dihydrofolate Reductase. *J. Am. Chem. Soc.* **2000,** *122* (2), 225-231.

29.     Guo, H.; Cui, Q.; Lipscomb, W. N.; Karplus, M., Substrate conformational transitions in the active site of chorismate mutase: Their role in the catalytic mechanism. *Proc. Natl. Acad. Sci. U. S. A.* **2001,** *98* (16), 9032-9037.

30.     Guo, H.; Cui, Q.; Lipscomb, W. N.; Karplus, M., Understanding the role of active-site residues in chorismate mutase catalysis from molecular-dynamics simulations. *Angewandte Chemie (International ed. in English)* **2003,** *42*, 1508-11.

31.     Martí, S.; Andrés, J.; Moliner, V.; Silla, E.; Tuñón, I.; Bertrán, J., Preorganization and reorganization as related factors in enzyme catalysis: the chorismate mutase case. *Chemistry-A European Journal* **2003,** *9* (4), 984-991.

32.     Copley, S. D.; Knowles, J. R., The conformational equilibrium of chorismate in solution: implications for the mechanism of the non-enzymic and the enzyme-catalyzed rearrangement of chorismate to prephenate. *J. Am. Chem. Soc.* **1987,** *109* (16), 5008-5013.

33.     Madurga, S.; Vilaseca, E., SCRF study of the conformational equilibrium of chorismate in water. *Phys. Chem. Chem. Phys.* **2001,** *3* (17), 3548-3554.

34.     Martí, S.; Andrés, J.; Moliner, V.; Silla, E.; Tuñón, I.; Bertrán, J., Conformational equilibrium of chorismate. A QM/MM theoretical study combining statistical simulations and geometry optimisations in gas phase and in aqueous solution. *J. Mol. Sturc-THEOCHEM* **2003,** *632*, 197-206.

35.     Elstner, M.; Porezag, D.; Jungnickel, G.; Elsner, J.; Haugk, M.; Frauenheim, T.; Suhai, S.; Seifert, G., Self-consistent-charge density-functional tight-binding method for simulations of complex materials properties. *Physical Review B* **1998,** *58* (11), 7260-7268.

36.     Elstner, M.; Frauenheim, T.; Kaxiras, E.; Seifert, G.; Suhai, S., A Self-Consistent Charge Density-Functional Based Tight-Binding Scheme for Large Biomolecules. *physica status solidi (b)* **2000,** *217* (1), 357-376.

37.     Schlitter, J., Estimation of absolute and relative entropies of macromolecules using the covariance matrix. *Chem. Phys. Lett.* **1993,** *215* (6), 617-621.

38.     Schäfer, H.; Mark, A. E.; van Gunsteren, W. F., Absolute entropies from molecular dynamics simulation trajectories. *J. Chem. Phys.* **2000,** *113* (18), 7809-7817.

39.     Schäfer, H.; Smith, L. J.; Mark, A. E.; van Gunsteren, W. F., Entropy calculations on the molten globule state of a protein: Side-chain entropies of α-lactalbumin. *Proteins: Struct. Funct. Bioinform.* **2002,** *46* (2), 215-224.

40.     Siegbahn, P. E. M., The performance of hybrid DFT for mechanisms involving transition metal complexes in enzymes. *J. Biol. Inorg. Chem.* **2006,** *11* (6), 695-701.

41.     Claeyssens, F.; Harvey, J. N.; Manby, F. R.; Mata, R. a.; Mulholland, A. J.; Ranaghan, K. E.; Schütz, M.; Thiel, S.; Thiel, W.; Werner, H.-J., High-Accuracy Computation of Reaction Barriers in Enzymes. *Angew. Chem.* **2006,** *118*, 7010-7013.

42.     Claeyssens, F.; Ranaghan, K. E.; Lawan, N.; Macrae, S. J.; Manby, F. R.; Harvey, J. N.; Mulholland, A. J., Analysis of chorismate mutase catalysis by QM/MM modelling of enzyme-catalysed and uncatalysed reactions. *Organic & biomolecular chemistry* **2011,** *9*, 1578-90.

43.     Chipot, C.; Pohorille, A., *Free energy calculations*. Springer: 2007.

44.     Olucha, J.; Ouellette, A. N.; Luo, Q.; Lamb, A. L., pH Dependence of Catalysis by Pseudomonas aeruginosa Isochorismate–Pyruvate Lyase: Implications for Transition State Stabilization and the Role of Lysine 42. *Biochemistry* **2011,** *50* (33), 7198-7207.

45.     Jorgensen, W. L.; Chandrasekhar, J.; Madura, J. D.; Impey, R. W.; Klein, M. L., Comparison of simple potential functions for simulating liquid water. *The Journal of Chemical Physics* **1983,** *79* (2), 926-935.

46. MacKerell, A. D.; Bashford, D.; Bellott, M.; Dunbrack, R. L.; Evanseck, J. D.; Field, M. J.; Fischer, S.; Gao, J.; Guo, H.; Ha, S.; Joseph-McCarthy, D.; Kuchnir, L.; Kuczera, K.; Lau, F. T. K.; Mattos, C.; Michnick, S.; Ngo, T.; Nguyen, D. T.; Prodhom, B.; Reiher, W. E.; Roux, B.; Schlenkrich, M.; Smith, J. C.; Stote, R.; Straub, J.; Watanabe, M.; Wiórkiewicz-Kuczera, J.; Yin, D.; Karplus, M., All-Atom Empirical Potential for Molecular Modeling and Dynamics Studies of Proteins. *J. Phys. Chem. B* **1998,** *102* (18), 3586-3616.

47. Mackerell, A. D.; Feig, M.; Brooks, C. L., Extending the treatment of backbone energetics in protein force fields: Limitations of gas-phase quantum mechanics in reproducing protein conformational distributions in molecular dynamics simulations. *J. Comput. Chem.* **2004,** *25* (11), 1400-1415.

48. Darden, T.; York, D.; Pedersen, L., Particle mesh Ewald: An N·log(N) method for Ewald sums in large systems. *J. Chem. Phys.* **1993,** *98* (12), 10089-10092.

49. Essmann, U.; Perera, L.; Berkowitz, M. L.; Darden, T.; Lee, H.; Pedersen, L. G., A smooth particle mesh Ewald method. *J. Chem. Phys.* **1995,** *103* (19), 8577-8593.

50. Tuckerman, M.; Berne, B. J.; Martyna, G. J., Reversible multiple time scale molecular dynamics. *J. Chem. Phys.* **1992,** *97* (3), 1990-2001.

51. Humphreys, D. D.; Friesner, R. A.; Berne, B. J., A multiple-time-step molecular dynamics algorithm for macromolecules. *The Journal of Physical Chemistry* **1994,** *98* (27), 6885-6892.

52. Berendsen, H. J. C.; Postma, J. P. M.; van Gunsteren, W. F.; DiNola, A.; Haak, J. R., Molecular dynamics with coupling to an external bath. *J. Chem. Phys.* **1984,** *81* (8), 3684-3690.

53. Hu, X.; Hu, H.; Yang, W. QM4D: An integrated and versatile quantum mechanical/molecular mechanical simulation package, http://www.qm4d.info/ (accessed 2016).

54. Ferrenberg, A. M.; Swendsen, R. H., Optimized Monte Carlo data analysis. *Phys. Rev. Lett.* **1989,** *63* (12), 1195-1198.

55. Kumar, S.; Rosenberg, J. M.; Bouzida, D.; Swendsen, R. H.; Kollman, P. A., THE weighted histogram analysis method for free-energy calculations on biomolecules. I. The method. *J. Comput. Chem.* **1992,** *13* (8), 1011-1021.

56. Hu, H., Wild-type and molten globular chorismate mutase achieve comparable catalytic rates using very different enthalpy/entropy compensations. *Sci. China. Chem.* **2014,** *57* (1), 156-164.

57. Zhang, X.; Bruice, T. C., Temperature Dependence of the Structure of the Substrate and Active Site of the Thermus thermophilus Chorismate Mutase E·S Complex. *Biochemistry* **2006,** *45* (28), 8562-8567.

**Figure legends**

**Scheme 1.** Schematic figure illustrating rearrangement of chorismate into prephenate.

**Figure 1.** Dihedrals changes of the C3–O7–C8–C9 (black triangle) and C2–C3–O7–C8 (red square) as a function of time for the chorismate conformation in water (A) and in PchB (B).

**Figure 2.** (A) PMF along the change of the C3–O7–C8–C9 dihedral in water and in PchB; (B) dihedral change in the enzyme starting from the local minimum in the enzyme; (C) representative geometry of local minimum; and (D) geometry of global minimum in PchB.

**Figure 3.** PMF profiles as the reaction coordinates of dihedral C3-O7-C8-C9 and the chemical reaction coordinate RC.

**Figure 4.** PMF of the uncatalyzed and catalyzed reaction at 298 K obtained by using (A) B3LYP/6-31G* and (B) SCC-DFTB. PMF profiles calculated by SCC-DFTB are obtained from 20 individual simulations for each window.

**Figure 5.** Least-squares fitting of the Gibbs–Helmholtz equation for the (A) uncatalyzed and (B) catalyzed reactions. Error bars are estimated from 20 independent simulations.

**Figure 6.** Free energy and entropy changes for the substrate ordering process and chemical reaction process.

# Figures

Scheme 1. Schematic figure illustrating rearrangement of chorismate into prephenate.

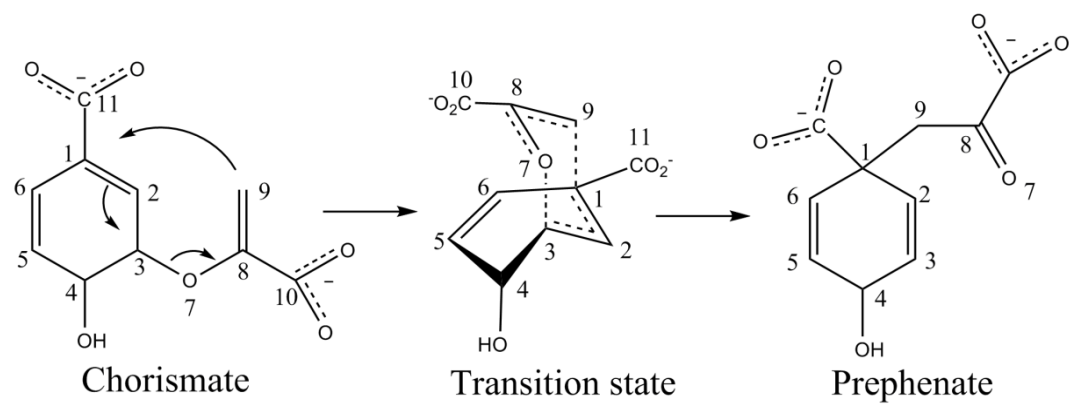

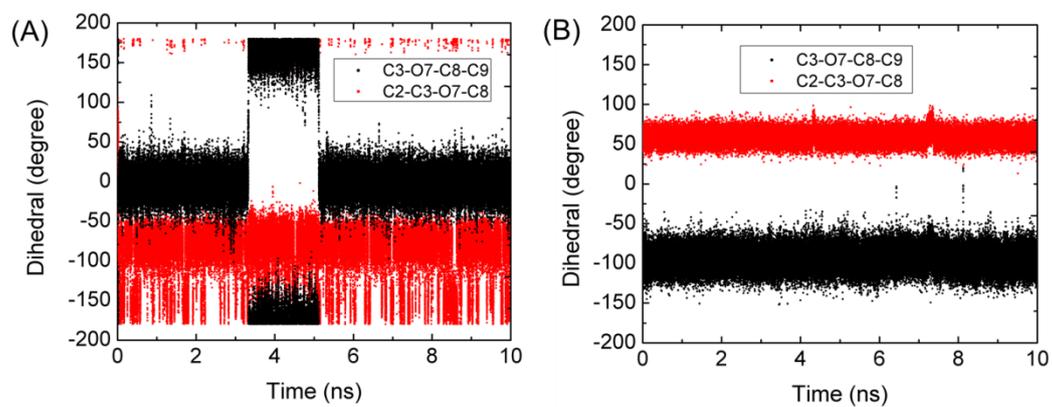

**Figure 1.** Dihedrals changes of the C3–O7–C8–C9 (black triangle) and C2–C3–O7–C8 (red square) as a function of time for the chorismate conformation in water (A) and in PchB (B).

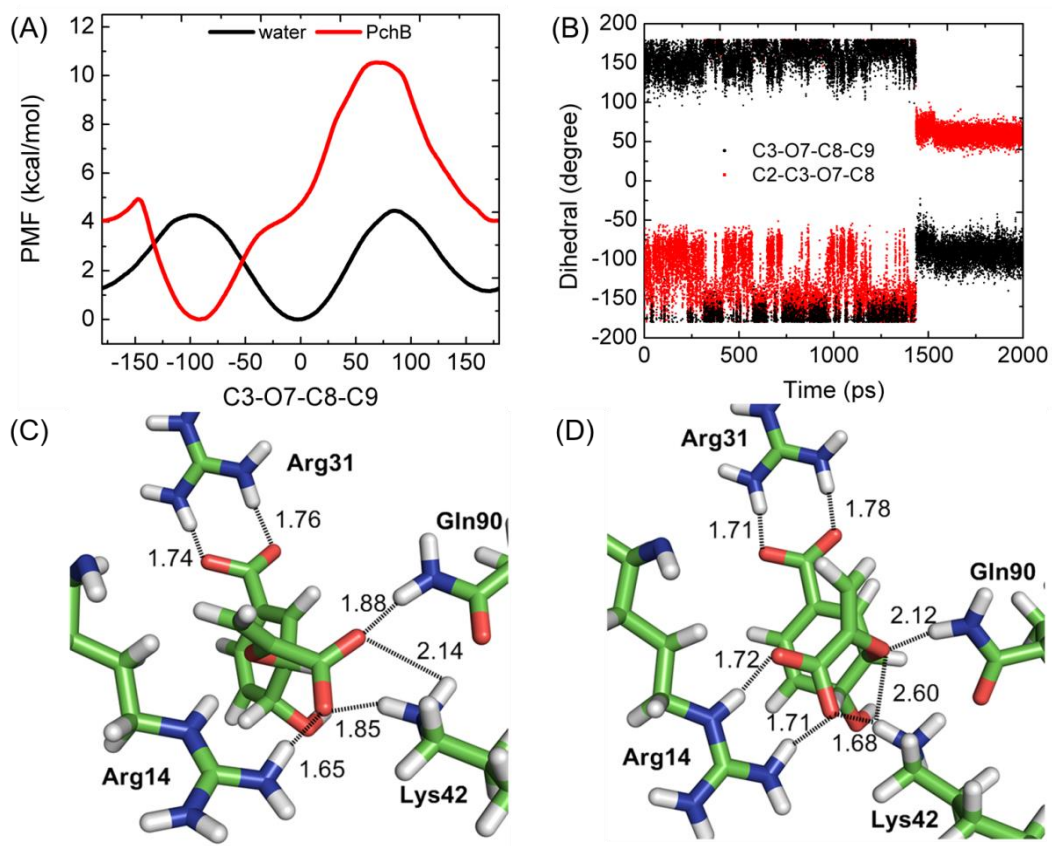

**Figure 2.** (A) PMF along the change of the C3–O7–C8–C9 dihedral in water and in PchB; (B) dihedral change in the enzyme starting from the local minimum in the enzyme; (C) representative geometry of local minimum; and (D) geometry of global minimum in PchB.

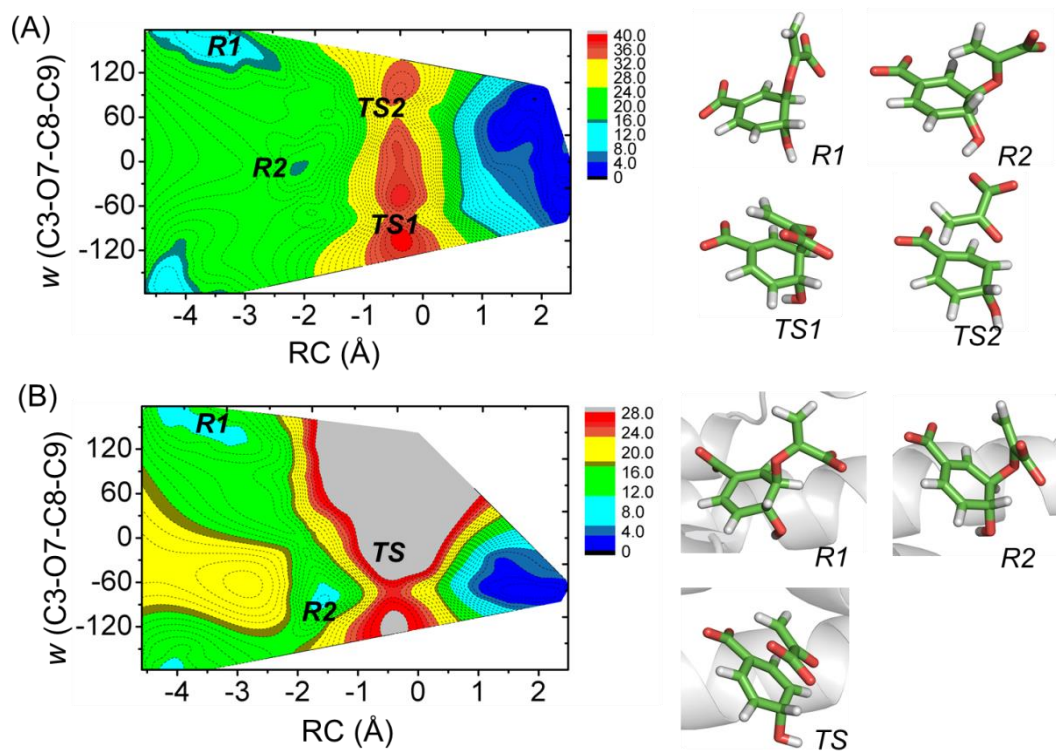

**Figure 3.** PMF profiles as the reaction coordinates of dihedral C3-O7-C8-C9 and the chemical reaction coordinate RC.

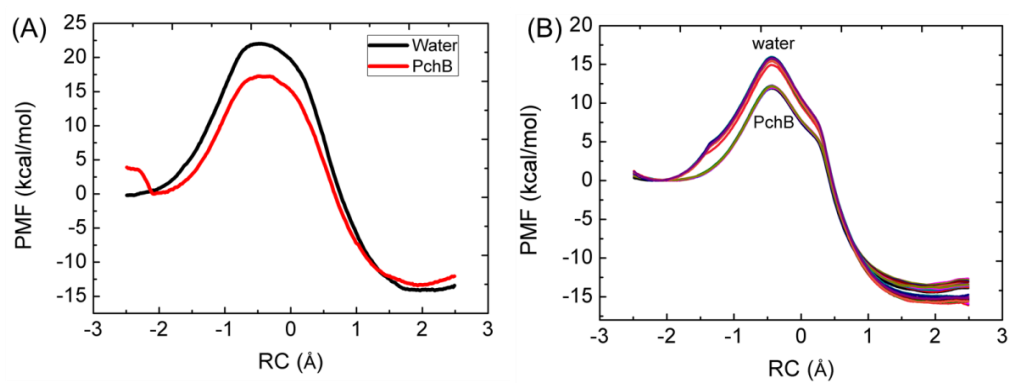

**Figure 4.** PMF of the uncatalyzed and catalyzed reaction at 298 K obtained by using (A) B3LYP/6-31G* and (B) SCC-DFTB. PMF profiles calculated by SCC-DFTB are obtained from 20 individual simulations for each window.

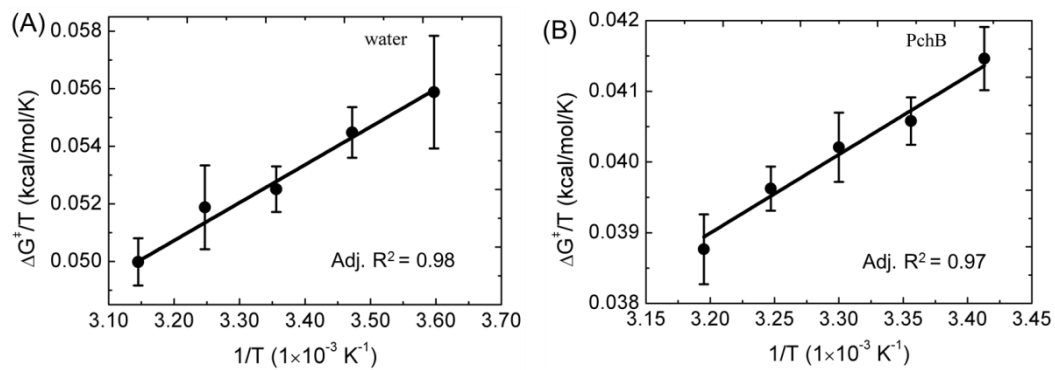

**Figure 5.** Least-squares fitting of the Gibbs–Helmholtz equation for the (A) uncatalyzed and (B) catalyzed reactions. Error bars are estimated from 20 independent simulations.

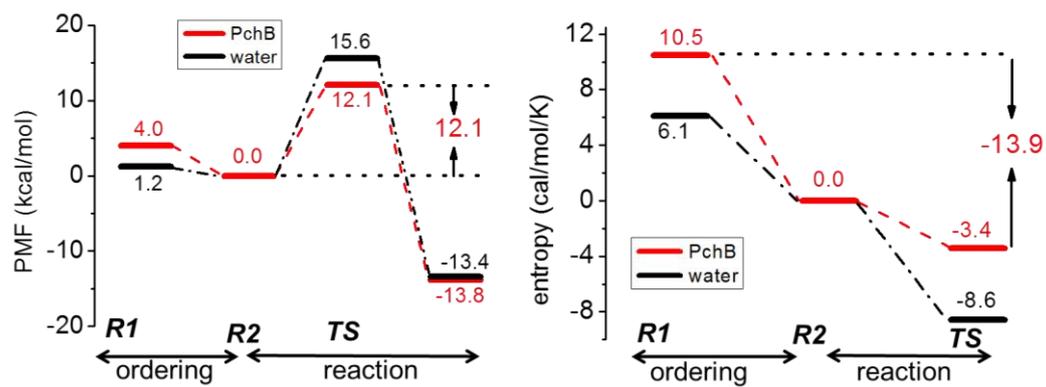

**Figure 6.** Free energy and entropy changes for the substrate ordering process and chemical reaction process.

# Tables

**Table 1.** Free energy, decomposed enthalpy and entropy for the reaction in PchB and in water.

|  | $\Delta G^{\ddagger}$ **(kcal/mol)** | $\Delta H^{\ddagger}$ **(kcal/mol)** | $-T(\Delta S^{\ddagger})$ **(cal/mol/K)** |
|---|---|---|---|
| **water** | 15.6(0.24) | 13.2 | 2.4 |
| **PchB** | 12.1(0.10) | 11.1 | 1.0 |
| **difference** | 3.5 | 2.1 | 1.4 |

**Table 2.** Key distance (in units of Å) and dihedral changes (in units of degree) for reactions in water and in PchB[a].

|  | water | | PchB | |
|---|---|---|---|---|
|  | *R2* | *TS* | *R2* | *TS* |
| **RC** | −2.11 | −0.45 | −1.97 | −0.43 |
| **d$_{C3-O7}$** | 1.49 | 1.85 | 1.51 | 1.91 |
| **d$_{C1-C9}$** | 3.60 | 2.30 | 3.48 | 2.34 |
| **d$_{C10-C11}$** | 6.48 | 4.84 | 5.23 | 4.73 |
| **w$_{C2-C3-O7-C8}$** | 57.4 | 54.9 | 56.9 | 53.9 |
| **w$_{C3-O7-C8-C9}$** | −4.9 | −73.1 | −95.7 | −74.3 |

[a]distances for the breaking bond (d$_{C3–O7}$), forming bond (d$_{C1–C9}$), and two carboxylate carbons (d$_{C10–C11}$); the dihedral describing the ether group with respect to the ring (w$_{C2-C3-O7-C8}$); and the dihedral describing the position of the side carboxylate group (w$_{C3-O7-C8-C9}$).

**Table 3.** Average distances (in units of Å) between enzyme residues and chorismate.

| chorismate–residue | PchB–*R2* | PchB–*TS* |
| --- | --- | --- |
| **O12-N1~Arg31** | 2.68 | 2.66 |
| **O13-N2~Arg31** | 2.74 | 2.76 |
| **O14-NE~Arg14(B)** | 2.70 | 2.72 |
| **O15-N1~Arg14(B)** | 2.69 | 2.66 |
| **O15-NZ~Lys42** | 2.68 | 2.71 |
| **O7-NZ~Lys42** | 3.12 | 2.99 |
| **O7-N~Gln90** | 2.98 | 2.92 |

**Table of Contents**

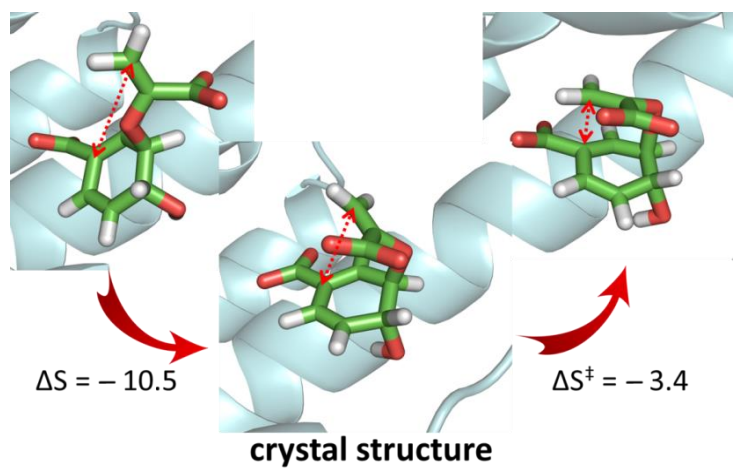